\documentclass{ieeeaccess}
\usepackage{cite}
\usepackage{amsmath,amssymb,amsfonts}
\usepackage{algorithm}
\usepackage{algorithmic}
\usepackage{graphicx}
\usepackage{caption,setspace}
\usepackage{textcomp}
\usepackage{comment}
\def\BibTeX{{\rm B\kern-.05em{\sc i\kern-.025em b}\kern-.08em
    T\kern-.1667em\lower.7ex\hbox{E}\kern-.125emX}}
\begin{document}
\history{Received 17 August 2023; revised 24 November 2023; accepted 11 December 2023; date of publication 15 December 2023; \\
date of current version 12 January 2024.}
\doi{10.1109/TQE.2023.3343625}

\title{Quantum Vulnerability Analysis to Guide Robust Quantum Computing System Design}
\author{\uppercase{Fang Qi}\authorrefmark{1}, 
\uppercase{Kaitlin N Smith}\authorrefmark{2}, 
\uppercase{Travis LeCompte}\authorrefmark{3},\\
\uppercase{Nian-feng Tzeng}\authorrefmark{4},
\IEEEmembership{Life Fellow, IEEE},
\uppercase{Xu Yuan}\authorrefmark{5},
\IEEEmembership{Senior Member, IEEE},\\
\uppercase{Frederic T Chong}\authorrefmark{6},
\IEEEmembership{Fellow, IEEE}, \\
and \uppercase{Lu Peng}\authorrefmark{1},
\IEEEmembership{Senior Member, IEEE}}

\address[1]{Tulane University, New Orleans, LA 70118, USA (email: \{fqi2, lpeng3\}@tulane.edu)}
\address[2]{Northwestern University, Evanston, IL 60208, USA (email: kns@northwestern.edu)}
\address[3]{Louisiana State University, 
Baton Rouge, LA 70803, USA (email: tlecom3@lsu.edu)}
\address[4]{University of Louisiana at Lafayette, Lafayette, LA 70503, USA (email: nianfeng.tzeng@louisiana.edu)}
\address[5]{University of Delaware, Newark, DE 19716, USA (email: xyuan@udel.edu)}
\address[6]{University of Chicago, Chicago, IL 60637, USA (email: chong@cs.uchicago.edu)}

\tfootnote{The work of Travis LeCompte was supported by a Louisiana Board of Regents Graduate Fellowship. This work was supported in part by the National Science Foundation (NSF) under Grant OIA-2019511, in part by Enabling Practical-scale Quantum Computing (EPiQC), an NSF Expedition in Computing, under Award CCF-1730449, in part by Software-Tailored Architecture for Quantum (STAQ) under Award NSF Phy-1818914, in part by the US Department of Energy (DOE) Office of Advanced Scientific Computing Research, Accelerated Research for Quantum Computing Program, in part by the NSF Quantum Leap Challenge Institute for Hybrid Quantum Architectures and Networks under NSF Award 2016136, in part by the U.S. Department of Energy, Office of Science, National Quantum Information Science Research Centers, in part by the Army Research Office under Grant W911NF-23-1-0077, and in part by the Oak Ridge Leadership Computing Facility, which is a DOE Office of Science User Facility supported under Contract DE-AC05-00OR22725.}

\markboth
{Qi \headeretal: QUANTUM VULNERABILITY ANALYSIS TO GUIDE ROBUST QUANTUM COMPUTING SYSTEM DESIGN}
{Qi \headeretal: QUANTUM VULNERABILITY ANALYSIS TO GUIDE ROBUST QUANTUM COMPUTING SYSTEM DESIGN}

\corresp{Corresponding author: Lu Peng (email: lpeng3@tulane.edu).}

\begin{abstract}
While quantum computers provide exciting opportunities for information processing, they
currently suffer from noise during computation that is not fully understood. Incomplete noise models have
led to discrepancies between quantum program success rate (SR) estimates and actual machine outcomes.
For example, the estimated probability of success (ESP) is the state-of-the-art metric used to gauge quantum
program performance. The ESP suffers poor prediction since it fails to account for the unique combination of
circuit structure, quantum state, and quantum computer properties specific to each program execution. Thus,
an urgent need exists for a systematic approach that can elucidate various noise impacts and accurately
and robustly predict quantum computer success rates, emphasizing application and device scaling. In this
article, we propose quantum vulnerability analysis (QVA) to systematically quantify the error impact on
quantum applications and address the gap between current success rate (SR) estimators and real quantum
computer results. The QVA determines the cumulative quantum vulnerability (CQV) of the target quantum computation, which quantifies the quantum error impact based on the entire algorithm applied to the
target quantum machine. By evaluating the CQV with well-known benchmarks on three 27-qubit quantum
computers, the CQV success estimation outperforms the estimated probability of success state-of-the-art
prediction technique by achieving on average six times less relative prediction error, with best cases at 30
times, for benchmarks with a real SR rate above 0.1\%. Direct application of QVA has been provided that
helps researchers choose a promising compiling strategy at compile time.
\end{abstract}


\titlepgskip=-15pt

\maketitle

\section{Introduction}
Excitement surrounds quantum computation due to the great theoretical potential both fault-tolerant~\cite{shor1999polynomial,grover1996fast} and near-term~\cite{moll2018quantum} quantum computers have to solve high-impact problems. By carefully leveraging quantum superposition, interference, and entanglement, quantum computers are projected to be applied to computational tasks that are currently intractable on today's most powerful supercomputers. Recent progress in quantum hardware has allowed many prototype quantum computers to emerge, and superconducting circuits \cite{clarke2008superconducting, dicarlo2009demonstration,stassi2020scalable} are gaining popularity as one of the forefront quantum computing technologies. Compared to other quantum hardware, superconducting quantum computers have advantages in scalability, microwave control, and nanosecond-scale gate operation \cite{fu2017experimental,fu2018microarchitecture,ding2020systematic}.

\begin{figure}[!t]
    \centerline{\includegraphics[width=0.5\textwidth]{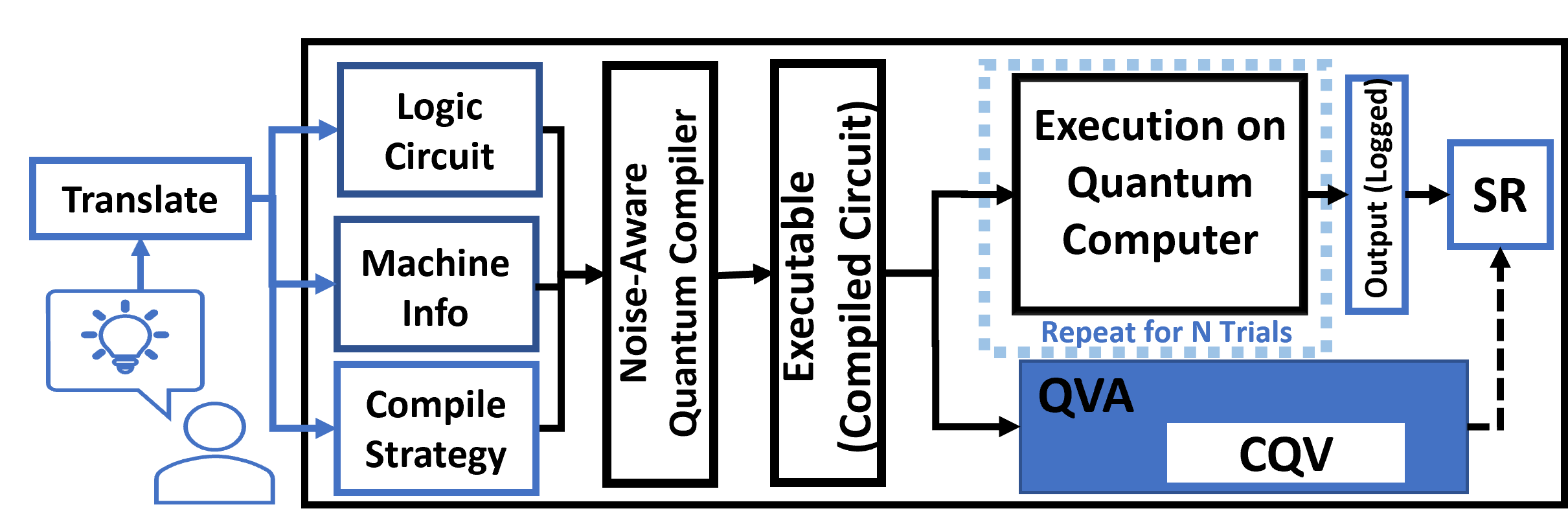}}
    \caption{Quantum experiment workflow with QVA. One complete cycle ranges from idea conception to real machine run.}
    \label{WholeStruct}
\end{figure}

While promising, superconducting quantum architectures are currently too error-prone to support programs targeted for large-scale applications. 
Near-term superconducting quantum computers suffer from various noise channels that degrade both quantum information and computation. This noise is difficult to fully characterize and causes retention and operational errors that vary both across-chip and between quantum computers\cite{tannu2019not} significantly. Quantum error correction was developed to accommodate occasional errors during quantum computation, but current noisy intermediate-scale quantum (NISQ) era machines do not have the operator precision or device scale to implement this routine~\cite {preskill2018quantum}. Therefore, NISQ quantum machines perform noisy operations as errors can happen on any physical qubit at any time during program execution according to error rates characterized by randomized benchmarking. Fig. \ref{WholeStruct} shows the full quantum computing flow that transforms a research idea into an experiment on a real quantum computer. 

Improving circuit success rate (SR) is a popular research topic, but few studies examine the quantum computer and circuit-dependent errors that result in specific SR. A robust noise model is essential for accurate SR prediction and developing SR boost technologies such as error reduction, bypass, and compression. Unfortunately, existing methods like statistical fault injection \cite{resch2020day, Qiskit, oliveiray2023systematic} and estimated probability of success (ESP) \cite{tannu2019ensemble} have accuracy and scaling challenges. These models neglect the impact of circuit composition on reported error rates from randomized benchmarking. Therefore, this article proposes a systematic approach to explain different errors and provide accurate SR prediction.

While studying ESP, we found that there are gates whose impact on the output is much more significant than their error rates.
We propose the quantum vulnerability analysis (QVA) to address these gate error inconsistent behaviors. 
QVA is a systematic methodology that performs error modeling based on randomized benchmarking to determine the success of a compiled circuit on a targeted quantum computer: the vulnerability metric cumulative quantum vulnerability (CQV) quantifies the compiled circuit performance, and 1-CQV provides an SR estimation that closely models actual quantum computer performance. By performing the QVA during compilation, researchers will have a clear and direct view of how gate error behaviors couple to the circuit structure to influence runtime performance. The blue box in Fig. \ref{WholeStruct} reveals the brief structure of the QVA and how CQV would be used in an experimental flow to estimate the quantum computer SR before real machine evaluation.

We extensively validated the accuracy and stability of QVA in predicting the SR of any given compiled circuits. This assertion is backed by over 160 K experiments conducted on three state-of-the-art 27-qubit IBM quantum computers~\cite{IBMQS}, spanning six distinguished algorithms with varying qubit sizes. The compiled circuits were produced using a variety of compilation strategies and incorporated multiple error-mitigation techniques to address the diverse noise profiles encountered over months of experimentation. This comprehensive approach bolsters the extendibility of our analysis to a broader range of circuits.

All results show that QVA maintains a stable SR estimation via 1-CQV, providing on average 6x improvements (30x in the best case) in relative prediction error over the widely implemented ESP estimator. Additionally, we conduct a case study to provide the quantum community with a direct application of the presented method by choosing promising compiling strategies at compile time. Further, our QVA module provides instructions for reconstructing the model based on a particular quantum device's topology and error behavior. It ensures its compatibility with various superconducting quantum computers, irrespective of their vendor or technology. Below are the contributions of our article:
\begin{itemize}
    \item We design and build a lightweight error modeling scheme based on QVA and are the first that quantify error rate impact propagating across the CNOT gates with the error rate reported by randomized benchmarking.
    \item We implement and evaluate a framework that calculates the unique CQV for an algorithm/machine pairing. We show that the proposed SR estimator, 1-CQV, outperforms the current state-of-the-art SR estimation, ESP, by an average 6x less relative prediction error.
    \item We highlight the scaling potential of CQV: as an algorithm reaches and surpasses the quantum volume of a device, CQV-based methods experience more than 10x improvement in relative prediction error rate compared with the state of the art.
\end{itemize}

\section{Background AND Related Work }
\subsection{NISQ Era Quantum Basics And Error Characterization}
The flow for generating a quantum executable from a given algorithm and evaluating it on a quantum chip is illustrated in Fig. \ref{WholeStruct}. 
The quantum compiler will be given information about the target quantum chip and compiler strategy, such as optimization levels, initial layout method, mapping method, etc. Based on that input, the compiler will follow all intermediate compiling steps to generate a compiled circuit for execution on a quantum computer. 

When operating a superconducting quantum computer, operations may fail due to poor environmental conditions, inaccurate control, state decoherence, and more. The error rate associated with an operation, \textbf{Operational Error Rates},  is closely estimated by randomized benchmarking, described in detail below, which approximates the extent of failures without revealing the exact source. The lifetime of a qubit, or its ability to retain a quantum state, is determined by its relaxation time (T1) and decoherence time (T2), so-called \textbf{Retention Errors}. The decoherence and relaxation time represent the qubit's average time to retain its energized and superimposed states, respectively.

\textbf{Randomized Benchmarking:} 
Operator performance must be accurately characterized to use a quantum computer effectively. Unfortunately, quantum computer noise models are complex, and it is unscalable to completely characterize system noise via process tomography~\cite{poyatos1997complete}. In addition to scaling considerations, characterization procedures must separate noise associated with quantum gates from errors stemming from state preparation and measurement to ensure that computation quality can be adequately estimated. 
Randomized benchmarking\cite{magesan2012characterizing,magesan3639robust} is a method of assessing quantum computer hardware that achieves an average error rate for operations through a process known as twirling. 
At the high level, twirling implements long sequences of random gate operations and fits the resulting data to a curve to determine the average error. 
Because randomized benchmarking considers only the exponential decay of sequences of random gates, sensitivity to measurement noise in the resulting average error is minimized. Meanwhile, the T1 and T2 errors on gate operation are included as part of the gate's error rate reported by the randomized benchmarking \cite{nishio2020extracting}.
Randomized benchmarking, while applicable to systems of any dimension, is predominantly employed for single-qubit or two-qubit gates~\cite{gaebler2012randomized}. This preference arises from the exponential growth in the number of required gates as the system's dimension increases, making the method less practical for larger systems. Despite this, the technique can be adapted to pinpoint errors due to unintended crosstalk~\cite{gambetta2012characterization}. Notably, IBM utilizes randomized benchmarking in each calibration cycle to ascertain error rates for their quantum computer's single and two-qubit gates, as reflected in the system's properties.

\begin{equation}
    Success\ Rate(SR) = \frac{Trial\ counts\ of\ correct\ output }{Total\  trail\ counts\ }\
    \label{SR}
\end{equation}

\textbf{Success Rate:} The success rate (SR) is used to gauge quantum program performance on a quantum computer. 
We compute SR by dividing the number of correct outputs by total executions, shown in Equation \ref{SR}. 
For more details on quantum computing, we refer to \cite{nielsen2002quantum}.

\subsection{Related Work on Quantum SR Estimation}
Early quantum computing research was focused on designing quantum hardware~\cite{majer2007coupling}, instruction set architecture \cite{fu2019eqasm}, and quantum computer microarchitecture \cite{fu2017experimental,fu2018microarchitecture, murali2020architecting, murali2019full}. 
Afterward, the temporal and spatial noise variation challenges of SC quantum computers were studied to discover mapping and allocation-enhanced compilations to make algorithm execution more robust to diverse errors\cite{tannu2019not, murali2019noise, lecompte2020robust, lecompte2023gnn, lecompte2023machine}.
Additionally, works such as \cite{kechedzhi2023effective} contribute to the understanding of how noise, fidelity, and computational cost interplay in quantum processing, enriching the broader discourse on quantum system performance.
Currently, the focus of quantum computing is on optimizing the success rate by applying different compiler strategies, such as mitigating the effect of errors by enhancing the quantum instructions \cite{shi2019optimized, gokhale2020optimized}, decreasing measurement errors  \cite{tannu2019mitigating,gokhale2020optimization}, mitigating crosstalk errors \cite{murali2020software, ding2020systematic}, combining pre- and post-execution software approaches to improve performance \cite{tannu2019ensemble, larose2022mitiq}, and compiling with specific constraints \cite{li2019tackling,ding2020systematic}.

While many studies have improved quantum program performance, two areas have been underexplored: 1) accurate SR prediction for specific compiled circuits and 2) better modeling of error/algorithm relationships in current quantum systems. 
Regarding SR prediction methods, popular alternatives to noisy quantum computer simulations include using machine learning for SR prediction, which treats the entire computation as a black box \cite{liu2020reliability}, developing detailed noise models, and methods like statistical fault injection \cite{Ibmquantum,oliveiray2023systematic, resch2020day} and estimated success probability (ESP) \cite{tannu2019ensemble, tannu2019not,nishio2020extracting}.

\begin{equation}
    g_{i}^{s}= (1- g_{i}^{e}),\   m_{i}^{s}= (1- m_{i}^{e})
    \label{errortosr}
\end{equation}

\begin{equation}
    ESP =  \displaystyle \prod_{i=1}^{N_{gates}} g_{i}^{s} * \displaystyle \prod_{i=1}^{N_{Measurement}} m_{i}^{s}
    \label{ESPequaiton}
\end{equation}

\section{Motivation} \label{sect:relatedwork}
\subsection{Limitation of Current SR Estimator}
\textbf{Machine Learning-based:}
The Machine Learning-based Success Rate (SR) prediction method, referenced in \cite{liu2020reliability}, simplifies quantum computations into a black box model. This approach can blur distinctions between circuits with varied gate parameters and requires retraining when adapting to different quantum machine sizes. Data collection for larger machines is resource-intensive, especially given the method's requirement to gather data within a single calibration period. Furthermore, its validation, limited to specific compiled circuits, may not account for the complexities of larger machines or diverse error-mitigation strategies.

\textbf{Fault Injection-based:}
The statistical fault injection method employs classical computers to simulate full-state quantum computations. During this process, errors are systematically injected into each basis gate based on specific triggering probabilities~\cite{Qiskit}. 
A very recent work \cite{oliveiray2023systematic} employs fault injection methods to evaluate quantum vulnerabilities. This study proposes the use of quantum fault injection to scrutinize circuit vulnerabilities, with a specific emphasis on radiation-induced errors. 
We note, however, that the fault injection approach may face challenges when applied to large machines. This complexity is accentuated when sampling a broad spectrum of circuits that exhibit variations in qubit size, circuit depth, and concurrent fault counts, especially in the context of larger machines and intricate algorithms.
Moreover, the quantum vulnerability factor (QVF) metric proposed in \cite{oliveiray2023systematic}, can face challenges when assessing circuits reaching the quantum supremacy, typically seen in machines with 50+ qubits \cite{arute2019quantum}.
The QVF calculation hinges on the contrast function, which necessitates prior knowledge of P(A), the expected correct state. In the absence of this knowledge, the correct state must be determined through a noise-free simulation. Relying on classical computing for full-state quantum circuit simulation poses a significant challenge, as such methods approach the limits of classical computational capabilities.
Alternatively, our research introduces a different approach, formulating a noise model addressing 1- and 2-qubit gate errors, measured errors, and crosstalk errors. Our methodology is crafted with a focus on scalability.

\textbf{Estimated Success Probability-based:}
The estimated success probability (ESP), shown in Equation \ref{ESPequaiton}, predicts the correct output trial probability by multiplying the success rate, or fidelity, of each gate ($g_{i}^{s}$) and measurement ($m_{i}^{s}$) operations, generated by one minus the gate ($g_{i}^{e}$) and measurement ($m_{i}^{e}$) error rate in equation \ref{errortosr}. 
While ESP considers all circuit operations, the product treats all gate errors that contribute to the final SR estimation equally when some gate errors influence the final circuit outcome more or less than others.
As a basic demonstration of the inaccuracy of ESP modeling, the gate success products in Equation~\ref{ESPequaiton} commute, whereas most operations in quantum circuits are fixed in ordering~\cite{matsuo2019reducing}. 
Based on such position differences of the gates on the compiled circuits, gate errors will contribute differently based on their propagation path to the measurement, which influence the estimated success rate differently than their original gate error rates. The detailed analysis is presented in Sec.\ref{sect:someerrronotmatter}. 
On the other hand, the simplicity of the ESP metric has made it frequently applied in quantum compiler design and circuit optimization efforts as a method to predict quantum program success on quantum computers \cite{tannu2019ensemble,acharya2020lightweight,patel2020ureqa,acharya2021automated,murali2020architecting,nishio2020extracting,li2022paulihedral}. However, if a better SR estimator was available, the effectiveness of the aforementioned quantum computer optimizations could potentially experience significant improvements.

\begin{figure}[!t]
\centerline{\includegraphics[width=0.5\textwidth]{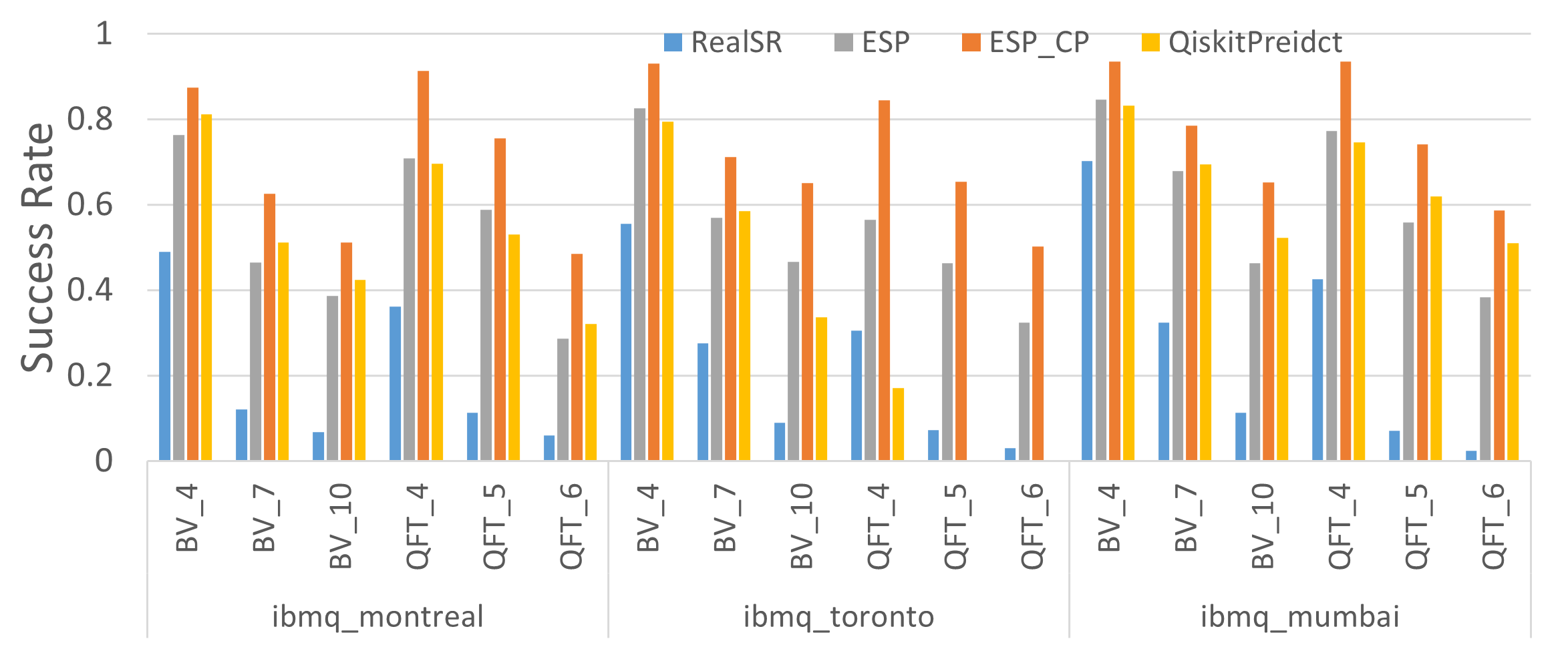}}
\caption{Current success rate estimators performance.}
\label{currentESP}
\end{figure}

We used statistical fault injection, ESP\_CP \cite{tannu2019not}, and ESP, as illustrated in Fig. \ref{currentESP}, to estimate the Success Rate (SR) of the Bernstein-Vazirani (BV) and Quantum Fourier Transform (QFT) algorithms across varying scales on three distinct quantum computers. The BV algorithm was selected due to its relatively shallow depth, allowing us to demonstrate the effects of significant algorithm size increments. Conversely, the QFT, characterized by deeper circuits, exhibits only modest size increases. For a comprehensive evaluation, both algorithms were tested with three different input sizes, yielding actual success rates ranging from 70\% to 5\%.
ESP\_CP, a variant of ESP, focuses solely on multiplying the success rates of gates situated on the critical path. 
Unfortunately, the predicted outcomes from both methods show significant deviation from the actual SR, with discrepancies ranging from 25\% to 60\% and relative error rates ranging between 70\% to 470\%. 
Further compounding the issue, both methods produce SR estimations that diverge sharply from the real machine results as the circuits increase in size, indicating that their tendency toward scaling is not well performed.

\begin{figure}[!t]
\centerline{\includegraphics[width=0.5\textwidth, height=1.2cm]{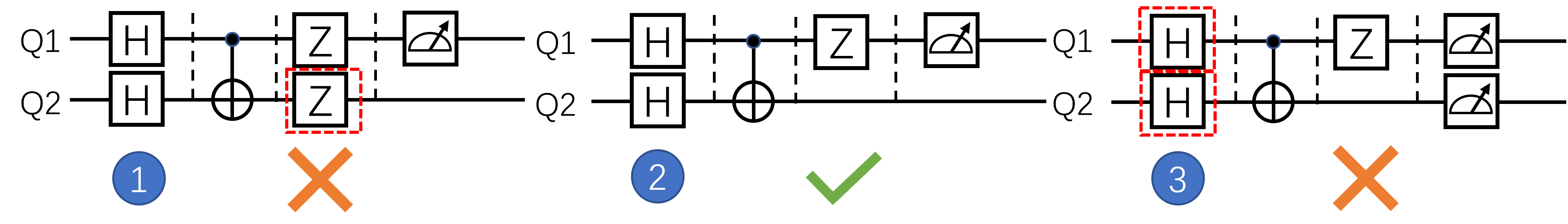}}
\caption{ESP for different circuits.}
\label{espissue}
\end{figure}

\subsection{Some Errors Matter, While Others Do Not} \label{sect:someerrronotmatter}
The ESP model, upon closer examination, presents potential sources of inaccuracies in SR prediction. Illustrated in Fig. \ref{espissue}, the ESP model accurately predicts the SR only for the middle circuit. When considering a compiled circuit, its final SR results from the product of individual qubit success rates, which are pre-measured. Therefore, only errors impacting a measurement gate influence the final output. 

For scenarios akin to the first circuit, the ESP model tends to overestimate error rates. Here, the error from the red-boxed $Z$ gate doesn't influence the measurement gate, meaning the $Z$ gate's error only impacts the success rate of $Q2$. Yet, this isn't captured by the subsequent $Q1$ measurement gate. 

Contrastingly, for situations resembling the third circuit, the ESP model tends to underestimate error rates. Errors originating from the two red-boxed $H$ gates not only affect the measurements of their associated qubits but also influence other qubits via the CNOT gate. Despite this, the gate error to success rate transformation, as outlined in equation \ref{errortosr}, only accounts for this error once. Any subsequent error impacts on other qubit measurements arising from error propagation are disregarded in the ESP model. 

Such observations underscore that certain errors exert more influence on the final output than others, highlighting the need for a refined approach. This analysis emphasizes the importance of taking into account circuit structure and architectural vulnerabilities when estimating SR on actual quantum computers.

\begin{figure}[!t]
\centerline{\includegraphics[width=0.5\textwidth]{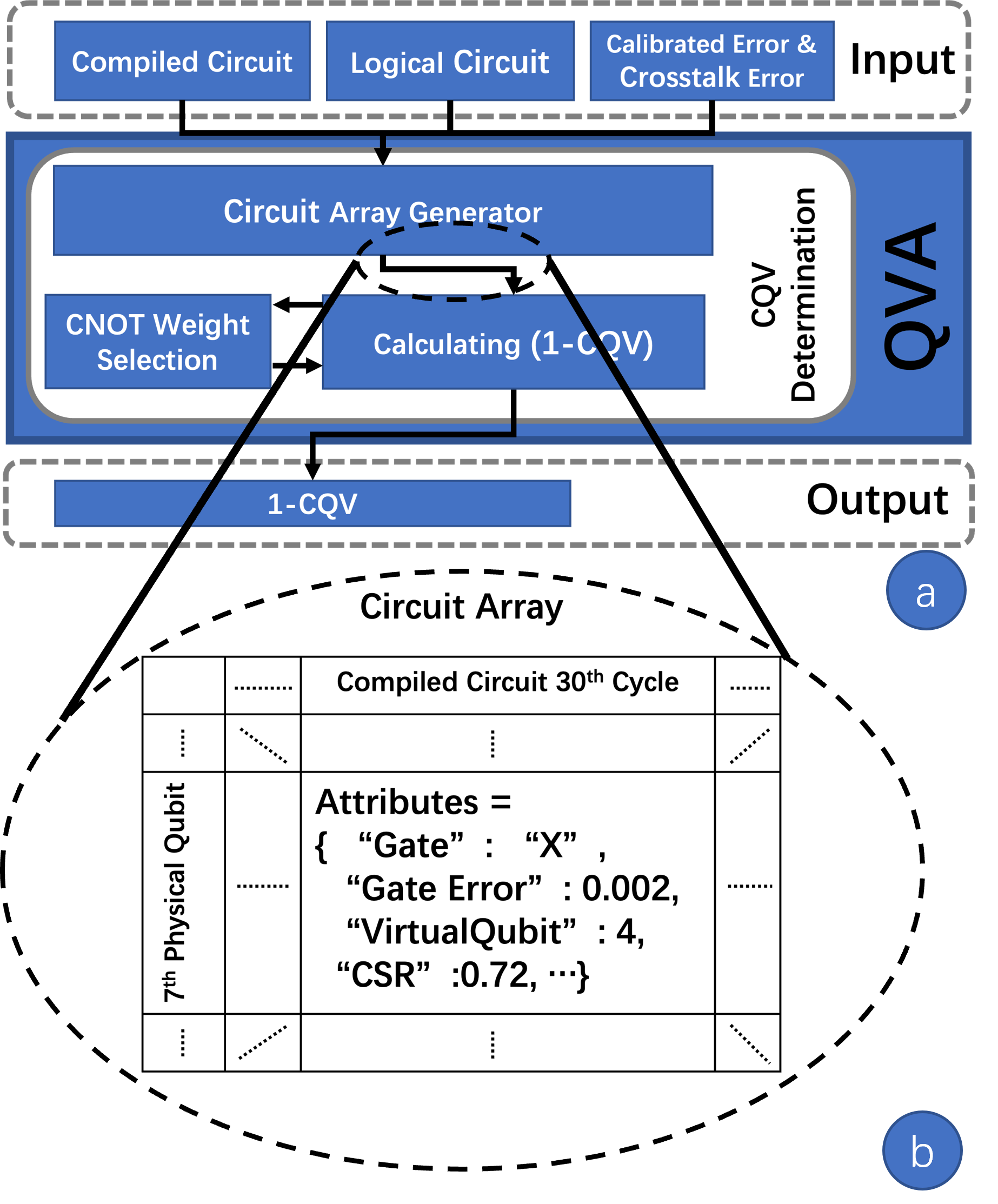}}
\caption{QVA workflow and circuit array example. }
\label{QVAFlow}
\end{figure}

\section{Quantum Vulnerability Analysis}
\subsection{QVA Overview}
In Section~\ref{sect:someerrronotmatter}, we discover that the SR of quantum computation is the outcome of the success rate of each qubit being measured. Each measured qubit's correctness is influenced by gate errors propagated to it. Our proposed quantum vulnerability analysis (QVA) is a systematic methodology that follows error propagation. The QVA will estimate the vulnerability of the compiled circuit by performing error modeling based on the error rates from randomized benchmarking calibration. 

The QVA generates the cumulative quantum vulnerability (CQV) metric. \textbf{Definition of CQV:} \emph{CQV presents the final circuit's vulnerability by predicting the failure rate (FR) for the compiled circuit.} We emphasize that the CQV will not predict the correct result but the possibility of an incorrect result during runtime. The calculation of $1-CQV$ represents the estimated success rate of a compiled circuit on the target machine calculated with the CQV. In Fig. \ref{QVAFlow}.a, we present the complete workflow of QVA. 

\subsection{CQV Determination} \label{sect: CQV }
\textbf{Circuit Array Generator: }
To understand the error propagation path within a quantum circuit during runtime, a connection between when an error occurs and how much it affects the compiled circuit must be established.
For more granularity, we quantify the compiled circuit to a finer degree by representing the algorithm at the cycle level. Cycle-level representation for a quantum circuit is analogous to the classical electrical circuit diagram to replace the previous analysis at the level of the complete compiled circuit. 
The Circuit Array Generator block transfers the compiled circuit further to a $2D$ array where each element represents the attributes of the physical qubit at that cycle, as shown in Fig. \ref{QVAFlow}.b. 
The circuit array records each physical qubit's attribution in every cycle, including the gate type, gate error, associated virtual qubit, its cumulative success rate, etc. 
Based on the cycle level compiled circuit, a snapshot of the operating quantum chip at a given cycle can be linked with the corresponding cycle in the compiled circuit.

\textbf{Calculating ($1-CQV$):}
The CQV methodology aims to predict the failure rate for a given compiled circuit on a designated quantum chip by effectively modeling the errors based on its propagation.  The Calculating $1-CQV$ block of Figure~\ref{QVAFlow}.a will first receive a circuit array with attributes filled. 
Next, we perform a crosstalk error calibration based on \cite{murali2020software} and update it to the $gate\ error$ by multiplying their success rate based on Equation \ref{errortosr}.  To determine the success rate estimation, which is \(1-CQV\), we introduce an algorithm (referenced as Algorithm \ref{algorithm}). This algorithm progressively updates the cumulative success rate (CSR) of each physical qubit based on the circuit array, considering the propagation of errors.

The algorithm initiates by setting the CSR for all entries in the Circuit Array to a perfect score, i.e., CSR = 1 (100\% success), as depicted in line 3. Subsequent steps, from lines 3 to 16, loop through all the gates, updating CSR values. For single-qubit gates, lines 6 and 7 modify the CSR for that gate by multiplying its total success rate with the preceding CSR value of the same qubit from the last cycle.

Complex operations like the CNOT gate necessitate a deeper understanding. Here, errors from one qubit can cascade to itself and affect the paired qubit. In lines 8 to 11, for every occurrence of such two-qubit interactions in a given compiled circuit, we introduce a weight \(w\). This weight, which lies between 0 and 1, signifies the fraction of cumulative error originating from the paired qubit that might propagate via the CNOT gate.

This error propagation model stems from the constraints imposed by the error rates disclosed through randomized benchmarking. The intricacy lies in the fact that these reported error rates cannot be disaggregated into individual types, such as phase, bit, or decoherence errors. Each type behaves differently when channeled through the CNOT gate.

For a target qubit involved in a CNOT gate, its CSR is computed as a product of its own success rate (\(g^s\)), its preceding CSR, and the success rate inherited from its paired qubit. This inherited success rate factors in only the weighted portion of the cumulative error. It's crucial to remember that the success rate (or the associated error rate) for any quantum element (qubit, gate, or circuit) can be deduced using equation \ref{errortosr}, which is based on the complementary relationship of success and error rates.

For idle cycles in the quantum circuit, we attribute an error value of zero. In the case of repeated gates, either compiler-introduced or manually inserted by the programmer, we appoint the error value based on the known error rates of such gates in the target machine.

In conclusion, the $1-CQV$ value, which represents the overall success rate prediction, is derived by multiplying the CSRs of all measurement gates. Upon examining Algorithm \ref{algorithm}, it becomes evident that its complexity is $O(G)$, where $G$ represents total gate counts for all types. This complexity arises because the algorithm iteratively checks each physical qubit's gate in every cycle to update the corresponding cumulative success rate. Consequently, the algorithm scales linearly with the gate count, encompassing all gate types, including ideal gates. This linear scalability of our noise model offers a significant advancement, facilitating both current NISQ-era and future quantum computing endeavors without being constrained by the limitations of classical computational power.

\textbf{CNOT Weight Selection:} In Sec. \ref{determineweight}, we have provided a study to observe the weight selection impact for various compiled circuits with different error profiles. Then we used a machine learning-based method to learn the weight value and used it to infer the proper weight in the CQV calculation. For more details. please refer to section \ref{determineweight}.

\begin{algorithm} 
\caption{Calculate $(1-CQV)$} \label{algorithm}
\begin{algorithmic}[1]
\renewcommand{\algorithmicrequire}{\textbf{Input:}}
\renewcommand{\algorithmicensure}{\textbf{Output:}}
\REQUIRE $Physical\ qubits\ QP$;$Compiled\ Circuit\ Cycles\ C$;
$Weight\ w$; $Circuit\ Array\  CA= [QP][C+1][Attr.]$; 
\ENSURE $(1-CQV)$ 
\STATE $(1-CQV) = 1$
\STATE let $Attr._{qp,c} = CA[qp][c][Attr.]$ for all Attributes
\STATE Initialize $CSR_{qp,0} = 1.00$ for every qubit at first cycle.
\FOR{each cycle $c$ from $1$ to $C+1$} 
\FOR{each physical qubit $qp$ in $QP$}

\IF{the $gate_{qp,c}$ is 1-qubit gate}
\STATE 	$CSR_{qp,c}$ = $g^s_{qp,c} * CSR_{qp,c-1}$ 
\ELSIF{the $gate_{qp,c}$ is CNOT gate }
\STATE 	$crosserror =(1- CSR_{qp^{'},c})* w$
\STATE 	$CSR_{qp,c} = g^s_{qp,c} * CSR_{qp,c-1} * (1-crosserror)$

\ENDIF
\ENDFOR
\IF{any $qp$ in the final swap cycle }
\STATE swap the $CSR_{qp,c}$ and $CSR_{qp^{'},c}$ for all swap pairs
\ENDIF
\ENDFOR
\STATE  $1-CQV =  \displaystyle \prod_{gate_{qp,c}= Measure} CSR_{qp,c} $
\RETURN $1-CQV$
\end{algorithmic}
\end{algorithm}

\begin{figure}[!t]
\centerline{\includegraphics[width=1.0\linewidth, height=3cm]{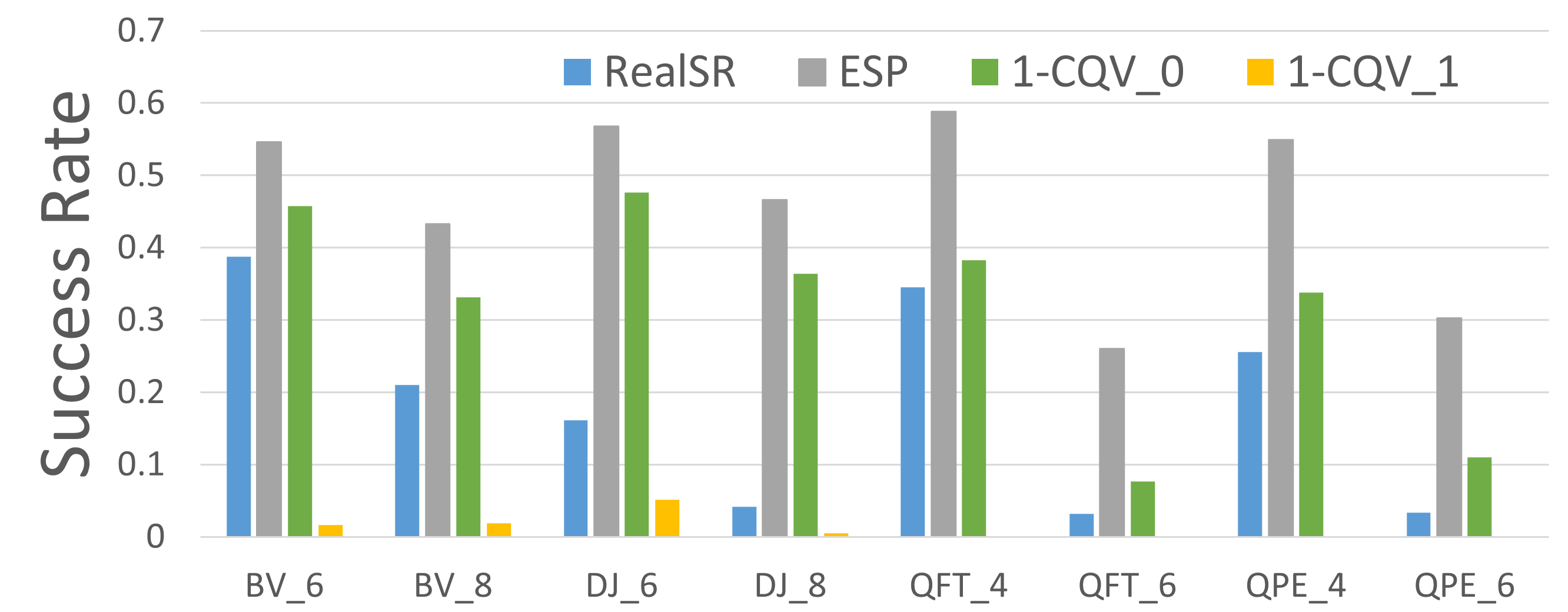}}
\caption{Average success rate prediction comparison between weight as 0 and 1 for benchmarks on IBMQ\_Montreal.  }
\label{figure: weightexist}
\end{figure}

\begin{figure}[!t]
\centerline{\includegraphics[width=1.0\linewidth]{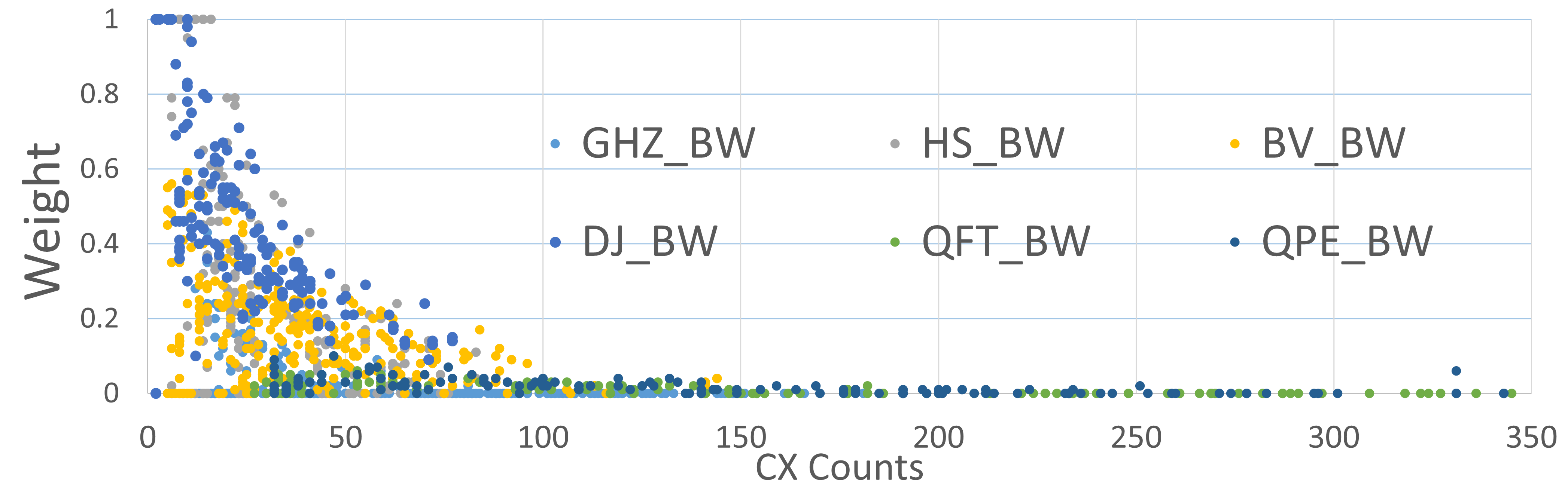}}
\caption{Comparing the best weight among all experiments.  }
\label{figure: weightdepth}
\end{figure}

\section{Determining CNOT Weight} \label{determineweight}
To accurately predict the real success rate, QVA requires a proper weight value, between zero and one, at compile time to assist the CQV calculation.
Before identifying the value of the weight, we first demonstrate the prediction performance when the weight is set equal to zero or one representing no error or full error crossover the CNOT gate, respectively.  The compiled circuits of the experiments are generated from different combinations of compiler settings for four benchmarks at two different algorithm sizes. The CQV calculation is performed using the calibration error and execution results from IBMQ\_Montreal on Apr. 1st, 2022. 
As shown in Fig. \ref{figure: weightexist}, though the CQV results with weight set to zero, $1-CQV_0$, are closer to the real success rate than ESP, there is still a non-trivial gap between $1-CQV_0$ and the real SR meaning that some errors are not well represented. 
Meanwhile, $1-CQV_1$ sets the weight to one, which makes the predictions close to zero all the time and loses track of the real SR, meaning the errors are being overestimated. 
The experimental results show that, for those benchmarks, using zero or one as the weight will lead to inaccurate predictions. 
When brute-force performing the CQV prediction for all the weights with 1\% granularity, we found the correlation between the weight value and its corresponding 1-CQV prediction is approximately -1. In other words, among all the weight values, there will always be one and only one weight value that returns a success rate prediction closest to the real SR, which will be labeled as the best weight. 

Based on such observation, we calculated the best weight for all compiled circuits generated from combining all the different algorithms, target machines, and compiled strategies. 
As shown in Fig. \ref{figure: weightdepth}, we have plotted the best weight against the CNOT count of the compiled circuit. The result is consistent with our expectation -- the best weight will be very arbitrary when the CNOT count is low, but as the CNOT count of the compiled circuit increases, the best weight begins to approach zero and shows an overall decreasing trend. 

\begin{figure}[!t]
\centerline{\includegraphics[width=1\linewidth]{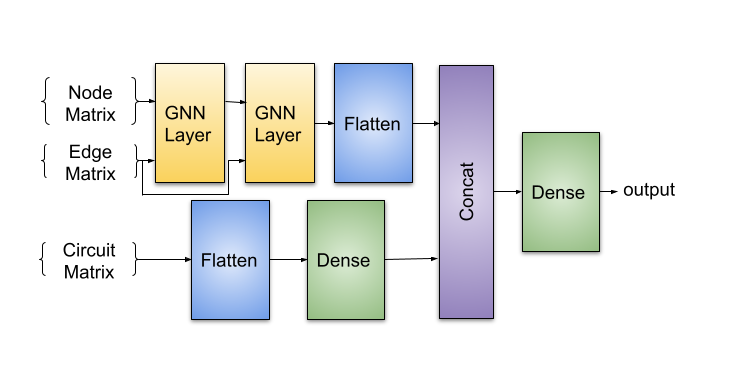}}
\caption{Graphic neural network-based model layout.}
\label{figure: Gnnmodel}
\end{figure}

\begin{figure}[!t]
\centerline{\includegraphics[width=1\linewidth]{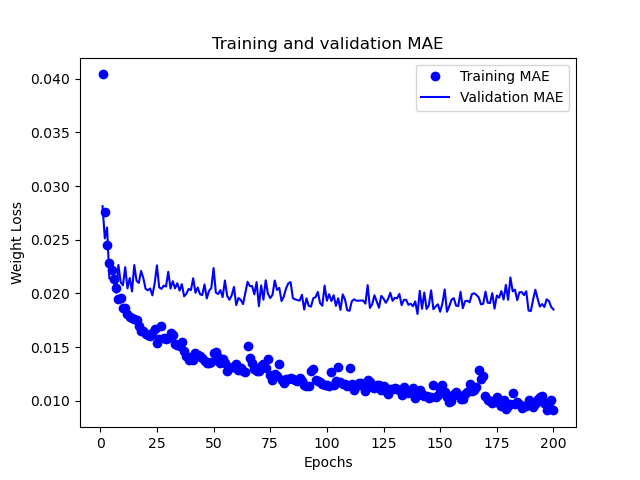}}
\caption{ GNN training performance for IBMQ\_Mumbai.}
\label{figure: GNNtrain}
\end{figure}

After analysis, we found that many factors, such as machine error properties, compiled circuit properties, etc., influence the best weight value. To fulfill the need of taking the graph-like machine information and circuit features into consideration, we choose Graph Neural Network (GNN) \cite{scarselli2008graph} and combine it with feed-forward networks to perform the best weight prediction shown in Fig. \ref{figure: Gnnmodel}. The Node Matrix represents the information for every physical qubit, including single-qubit operation error rates. The Edge Matrix in the figure presents the CNOT error rates for each physical qubit pair. The circuit matrix contains information for each qubit's operation count, measurement info, and two-qubit operations count. 
As shown in Fig. \ref{figure: GNNtrain}, by leveraging the GNN model, we can provide the weight value with an average 2\% difference to the best weight, which strongly supports an accurate SR prediction in later execution. The trained model will be used to infer the best weight for CQV calculation.  

\section{Implementation}
We employ Qiskit \cite{Qiskit}, a renowned open-source framework for quantum computing, as the foundation for implementing and assessing our QVA methodology. Our work extends Qiskit version 0.34.2, enabling it to execute QVA and compute the (1-CQV) for any specified compiled circuit.
Our QVA approach is meticulously crafted to provide an accurate and efficient estimation of the success rate for compiled circuits on specific quantum machines. To ensure a diverse set of compiled circuits, we utilize various combinations of compiling policies for each benchmark-machine pairing. This approach integrates a spectrum of error-mitigation techniques at every stage, ensuring comprehensive and robust evaluations. 
Table \ref{tab:algorihtms} describes the backends and benchmarks used in our evaluation. We repeatedly perform all the algorithms that range in scale from 4 input qubits to 15 input qubits over months, which generated 160K distinctive compiled circuits captured with diverse noise calibration profiles. To focus on meaningful results, we ignore experiments that return a real SR below 0.1\%. We chose the state-of-the-art SR estimator ESP for the baseline and ignored the ESP\_CP since it underestimates error. 

In the experimental flow, we first run the given compiled circuits on their target quantum computers for the default 8192 trials and log the outputs. Then, we use the Qiskit simulator to capture the correct result and generate the success rate of the experiment based on the logged output on the classical computer. 
For scenarios approaching quantum supremacy, where classical computers struggle with full-state quantum circuit simulations, we introduce an alternative method to ascertain correct results, as detailed in Sec. \ref{sec: overhead_disc}.
Now the experiment's real performance is known and named real SR. 
For each machine, we used the first ten days of data to perform the GNN training on weight selection and to infer the weight to assist CQV prediction for the rest of the experiment data. This offline training is a one-time procedure, and in our experiments, it took approximately an hour on an NVIDIA 3060 GPU. Then we perform ESP and 1-CQV, which estimate the real SR based on the calibrated error of the experiment on the classical computer. 
The estimated SR generated from ESP and 1-CQV will be compared with the real SR. In addition to quantifying the estimation accuracy directly by performing the absolute difference between the real and estimated SR, we also use the Relative Prediction Error metric, which uses the absolute difference divided by the real SR, to present the accuracy trend of the method while compensating for increases in the algorithm size resulting in decreasing real SR.

\begin{scriptsize}
\begin{table}[!h]
    \small
    \centering
    \caption{Benchmarks and Quantum Computer Description}
    \label{tab:algorihtms}
    \begin{tabular}{|l|l|}
    \hline 
    \textbf{Item} & \textbf{Description}\\
    \hline
    \hline
    \hline
     BV & Bernstein-Vazirani \\
    \hline
    DJ    & Deutsch-Jozsa \\
     \hline
    HS & Hidden Shift\\
     \hline
     GHZ & Greenberger-Horne-Zeilinger\\
     \hline
   QFT     &  Quantum Fourier Transform\\
    \hline
   QPE    &  Quantum Phase Estimation\\
    \hline
    \hline
    ibmq\_montreal        & 27-qubits with Hexagon \\
    \hline
    ibmq\_toronto          & 27-qubits with Hexagon  \\   
    \hline
    ibmq\_mumbai          & 27-qubits with Hexagon   \\   
    \hline
    \end{tabular}
\end{table}
\end{scriptsize}

\section{Results}

\begin{figure*}[!t]
    \centerline{\includegraphics[width=1.0\linewidth]{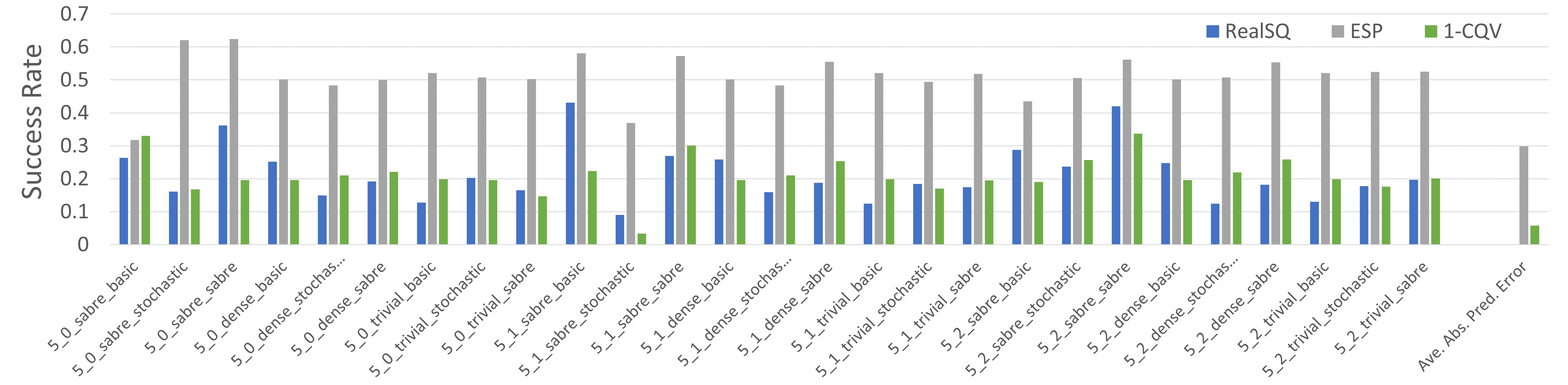}}
    \caption{Predicted and real success rates of QPE with IBMQ$\_$Montreal Quantum machine. For each compile configuration listed on the x-axis, the variables separated by an underscore are ordered as algorithm input qubits, optimization level, allocation method, and routing method.} 
    \label{fig：predict}
\end{figure*}

\subsection{The CQV Accuracy }
\textbf{Algorithm size within Quantum Volume:}
Here, we present the CQV prediction performance for all six benchmarks on the Quantum machines IBMQ\_Montreal, IBMQ\_Toronto, and IBMQ\_Mumbai.  
As shown in Fig. \ref{fig：predict}, we presented the 1-CQV prediction accuracy compared with ESP with varying compiled circuits for a five-qubit QPE algorithm on IBMQ\_Montreal on Apr. 5th, 2022. The different configurations are guidelines for the compiler to generate the final compiled circuit based on the given logical circuit and target device. 

We observe that $1-CQV$ is much closer to the real SR than the ESP. After being shown to the right of Fig.  \ref{fig：predict}, the average absolute error rate for ESP over all the configurations compared to ground truth SR is 29.8\%, while $1-CQV$ achieves an average error of 4.8\%. Such an error rate difference means that $1-CQV$ achieves an 84\% error reduction compared to the ESP, which also means the CQV calculation is adaptable to the variation of errors across different calibration periods and provides excellent predictions.

\begin{figure}[!t]
\centerline{\includegraphics[width=1.0\linewidth]{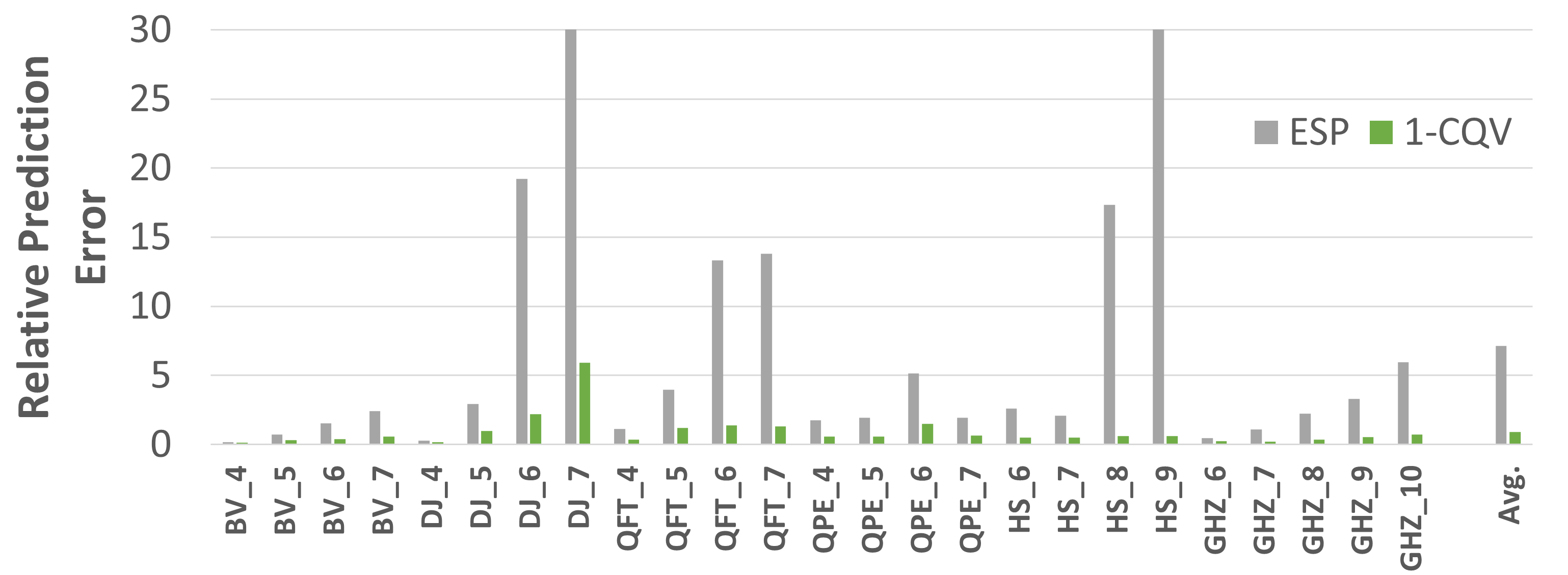}}
\caption{Average relative predict error for all the benchmarks on single Quantum machine IBMQ\_Montreal.}
\label{figure:singlemachine}
\end{figure}

\begin{figure}[!t]
\centerline{\includegraphics[width=1.0\linewidth]{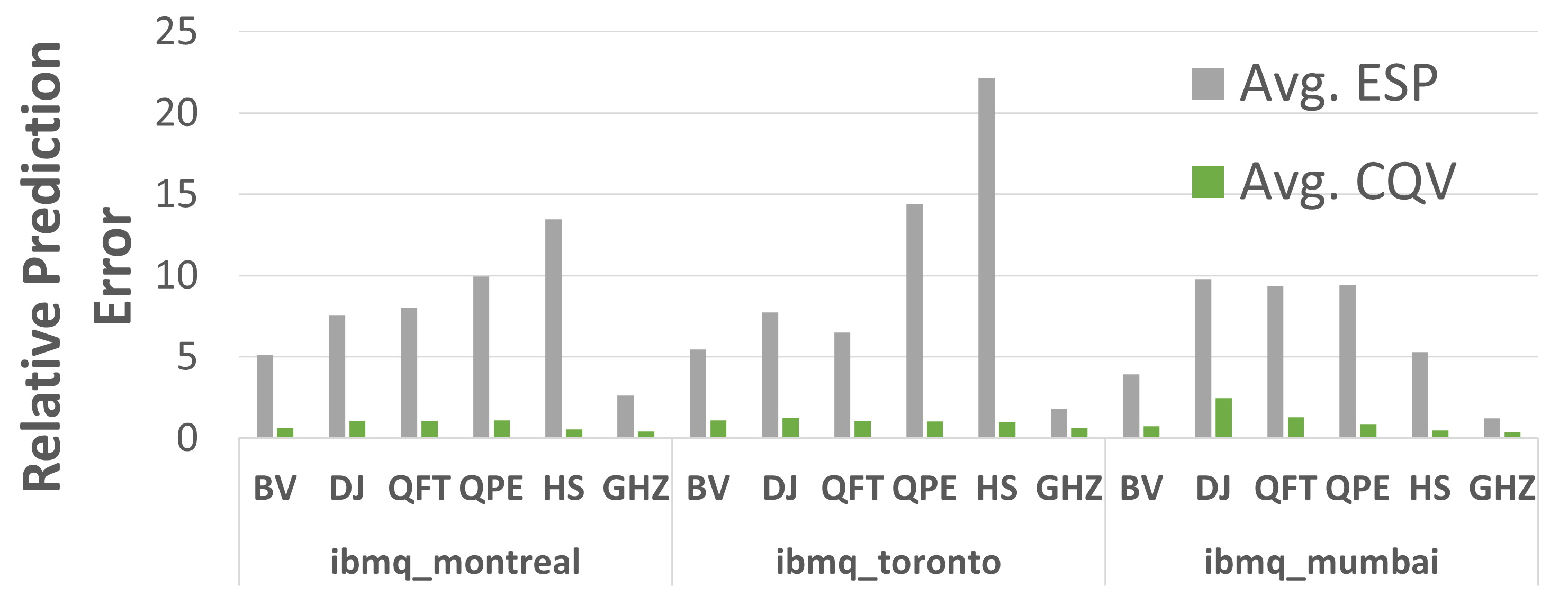}}
\caption{Average relative predict error across all benchmarks and backends.}
\label{allmachine}
\end{figure}

As shown in Fig. \ref{figure:singlemachine}, we present the 1-CQV prediction for all the benchmarks on the single machine IBMQ\_Montreal. We include experiments with an SR higher than 0.1\%. The relative prediction error is produced by dividing the absolute prediction error by the SR. The trend emerges that 1-CQV prediction outperforms the ESP by achieving an average of 6 times less relative prediction error rate and the best improvement of 30 times. Meanwhile, the relative prediction error jumps clearly when the algorithm size increases, which follows the nature of the benchmark by employing significantly more two-qubit gates. When increasing algorithm size, not only does the number of two-qubit gates increase, but the number of swapping operations also increases. As shown in Fig. \ref{allmachine}, 1-CQV presents a stable and accurate average relative prediction error across all machines and benchmarks. Based on the results, we conclude that the CQV achieved the goal of designing a more precise SR estimator consistently across different dates, machines, and algorithms than the state-of-the-art SR estimator.

\textbf{Algorithm size Beyond Quantum Volume:}
From the results shown in Fig. \ref{fig：predict}, \ref{figure:singlemachine}, \ref{allmachine}, $1-CQV$ proves to predict SR with a closer distance to the real SR even when it falls in the 10\%  to 0.1\% range. 
The primary reason for the low success rate is that the size of the compiled circuits equals or exceeds the desired quantum volume of the target quantum machine. Quantum volume can be defined as the product of the number of virtual qubits and maximum circuit depth supported by the machine. 

After making a full-spectrum comparison among all the backends and benchmarks, from the observations of the results, we can say that the CQV has better error modeling than the ESP noise model. Fig. \ref{figure:singlemachine} demonstrates that the CQV prediction is stable across different algorithms with a much lower relative error rate. Additionally, the results show that CQV performs much better when the algorithm reaches or exceeds the quantum volume, which is a valuable property when the limited quantum volume is the bottleneck of the current NISQ era.
Therefore, we conclude that the $1-CQV$ prediction is accurate across the full spectrum of algorithm sizes and success rates.

\begin{table}[!h]
    \centering
    \caption{Execution Time Comparison(Seconds) }
    \label{tab: scaling}
    \begin{tabular}{|l|l|l|l|l|}
    \hline 
    \textbf{Algorithm} &\textbf{CX count}& \textbf{ESP}& \textbf{Qiskit}& \textbf{CQV}\\
    \hline
    QFT\_5  &59  & 0.00299& 1.31142 & 0.01830\\
    \hline
    QFT\_10  & 408  &0.02004 & 1.86264  &0.05310 \\
    \hline
    QFT\_20   & 2657   &0.09802 & $>10$ mins  &  0.25227\\
    \hline
    QFT\_50   &   26408  &0.71115  & N/P & 6.01701\\
    \hline
    QFT\_100 &    157428   &   1.83842  & N/P &7.82150 \\   
    \hline
    QFT\_120   &    208260  &  3.10630 & N/P & 20.4157 \\   
    \hline
    \hline
    \end{tabular}
\end{table}

\subsection{Scalability Analysis of CQV } \label{sec: overhead_disc}
In the rapidly evolving landscape of quantum computing, the intricacy of quantum circuits is escalating, bringing us closer to addressing real-world challenges. As we navigate this frontier, the need for prediction models that are both accurate and scalable becomes paramount. Our study, therefore, focuses on the scalability of the Quantum Vulnerability Analysis (QVA), examining its efficiency across a spectrum of quantum circuit sizes.

The Quantum Fourier Transform (QFT) was chosen as the benchmark for our evaluation. With input sizes spanning from 5 to 120 qubits, it's noteworthy that machines operating with around $50$ qubits are on the threshold of quantum supremacy \cite{arute2019quantum}. Our scalability experiments for QVA were conducted based on the topology and error profile of the IBMQ\_Washington machine, a state-of-the-art 127-qubit quantum computer.

To ensure our execution time assessment was both transparent and unbiased, we considered only the inference time of the Graph Neural Network (GNN), which provides the CNOT weight value, and the execution time of the CQV noise model. This choice was made because the GNN training is an offline process, executed just once. As shown in Table \ref{tab: scaling}, our CQV approach demonstrated linear scalability, with execution time increasing proportionally with the CNOT gate counts. 
It is worth mentioning that the execution of the CQV takes place on the CPU, including the pre-trained GNN to infer the weight value $w$. 
This linear trajectory underscores the potential of our model to efficiently manage complex quantum circuits in the future.

However, the full-state quantum circuit simulator-based approaches face challenges in scalability, particularly with circuits that exceed 32 qubits. This limitation highlights the simulator's constraints when tasked with emulating large-scale, real-world quantum systems.
In contrast, our approach requires only a fraction of the computational power for estimating success rates. 
To both validate our predictions and refine our noise model, particularly for circuits approaching quantum supremacy, we've devised a novel method: by merging the original circuit with its inverse and using the input states as a benchmark, we can effectively measure the circuit's performance and fine-tune our noise model. 
It's pertinent to note potential overfitting for specific circuits due to the reliance on reverse circuits. 
Nevertheless, we believe our approach adds a valuable technique to the toolbox for exploring circuit vulnerabilities.

Furthermore, while our preliminary results on QVA's scalability are promising, they represent just the tip of the iceberg. Our model's design is inherently versatile, unencumbered by specific error rates, gate types, or circuit structures. This adaptability not only allows for the integration of partial error correction techniques in larger devices but also hints at the vast potential for future optimization strategies, further refining the success rate estimation process.

\section{Case Study: Choosing Compiling Strategy }
\begin{figure}[!t]
\centerline{\includegraphics[width=1.0\linewidth]{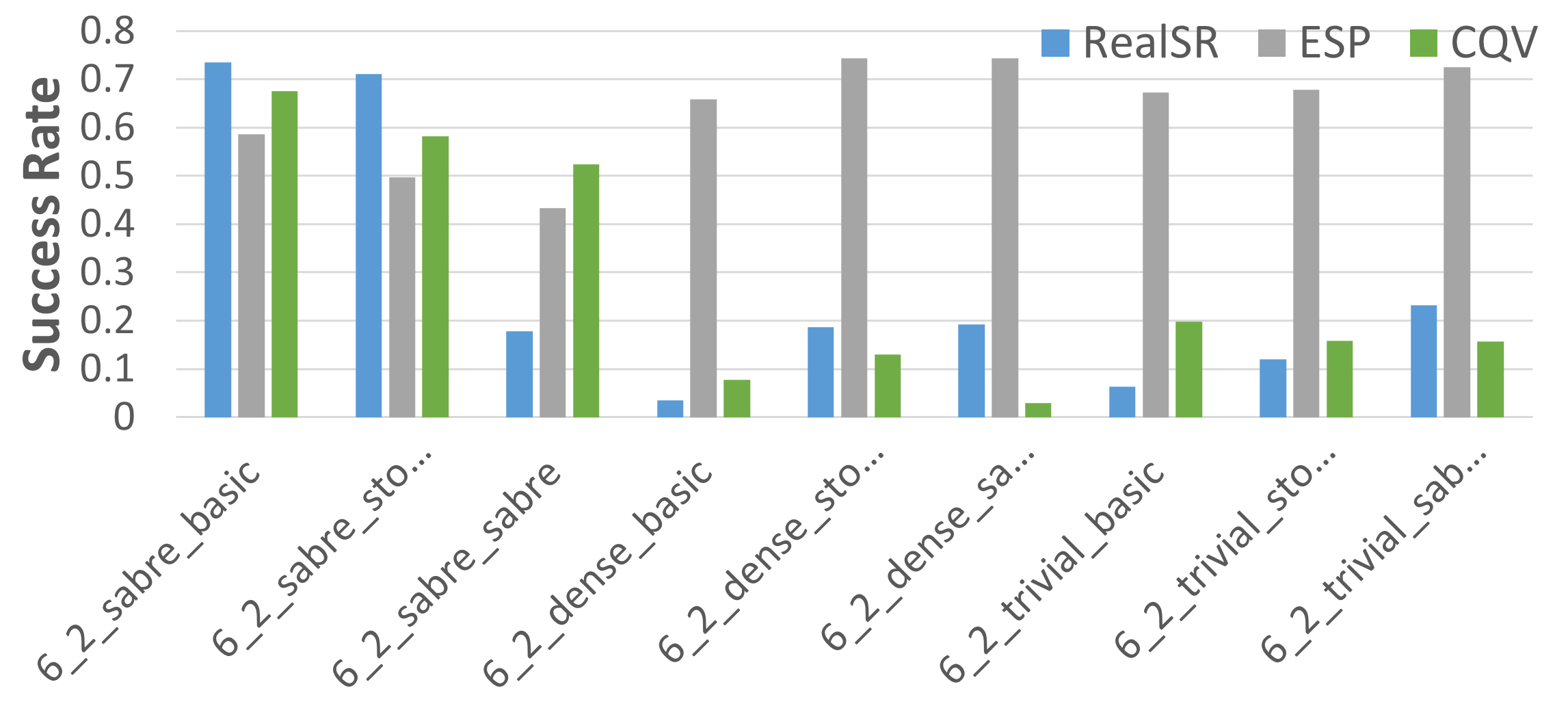}}
\caption{Choose compiler policies for HS on IBMQ\_Toronto.}
\label{compilerhelper}
\end{figure}

The current access modes for quantum computers are either limited free access to small machines or expensive hourly institutional subscriptions to large devices. Naturally, users will want the highest SR with as few executions as possible to save time, money, and access. 
However, finding the best combination among all the available machines and compiler configurations to achieve the best performance is challenging. The search space will grow enormously when also considering compilation optimizations. Without performing a brute-force execution of all the combinations, identifying the best strategy is challenging. Since CQV is more accurate than ESP, we would naturally ask, could CQV be used to suggest the compiling configuration with the optimal performance? 
To answer that, we performed both ESP and CQV for all the combinations of compiler configuration at compile time for the HS benchmark and picked the two configurations with the highest estimated SR for both ESP and CQV. Based on the result, the CQV outperforms the ESP across all algorithm sizes. One example of HS at level 2 optimization and six input states is shown in Fig. \ref{compilerhelper}. It is clear that the two choices with the highest 1-CQV estimation not only have less prediction error but also result in the highest Real SR. In contrast, the top two high-ranking ESP configurations result in the 4th and 5th best in real SR out of 9 combinations. Moreover, it will be the 7th and 8th choice of ESP to pick up the compiler configurations for the highest real SR.
For the computation overhead, executing CQV prediction is acceptable since the calculation is done on a classical computer. Furthermore, since the execution of all the SR predictions is independent of each other, it is possible to perform parallel computing for different compiler strategies, and the individual prediction overhead is discussed in \mbox{\ref{sec: overhead_disc}. }
In this case, the CQV can guide a user to choose a more effective and reliable compiler strategy with a higher SR than ESP. No prediction can be perfect (that would be computationally intractable), but CQV improves prediction enough to be usable for compiler decisions.

\section*{ACKNOWLEDGMENTS}
The views and conclusions contained in this document are those of the authors and should not be interpreted as representing the official policies, either expressed or implied, of the U.S. Government. The U.S. Government is authorized to reproduce and distribute reprints for Government purposes notwithstanding any copyright notation herein.

\section{Conclusion}
In the rapidly evolving field of quantum computing, predicting the success rate (SR) of a quantum circuit remains a challenging task. Existing methodologies often fall short, either by oversimplifying the error model or by not adequately accounting for the intricacies of error propagation within complex quantum circuits. Recognizing this gap, we present the Quantum Vulnerability Analysis (QVA) in this paper, a robust systematic approach to determining a given computation's Cumulative Quantum Vulnerability (CQV). The QVA offers a nuanced, detailed method to estimate the failure rate of a given compiled circuit, considering the effects of individual gate errors, their cumulative influence, and the unique properties of quantum gates such as CNOT. To establish the efficacy of QVA, we subjected it to rigorous validation on cutting-edge quantum machines using well-known benchmarks. The results demonstrated that QVA consistently outperformed the prevalent success rate estimator, ESP, and showcased linear scalability. On average, our model exhibited a six-fold reduction in the relative prediction error rate compared to ESP. Such accuracy not only bolsters confidence in our method but also has far-reaching implications at both the hardware and software levels.

\bibliographystyle{IEEEtranS}
\bibliography{tqe}

\EOD

\end{document}